\documentclass[review]{elsarticle}
\usepackage{graphicx}
\usepackage{amsmath}
\usepackage{enumerate}

\usepackage{natbib}
\setcitestyle{authoryear,open={(},close={)}}

\journal{----}

\graphicspath{Figures}             
\DeclareGraphicsExtensions{.pdf,.jpeg,.png,.eps}

\usepackage{subfigure}
\usepackage{multirow}
\usepackage{setspace}
\usepackage{epstopdf}
\usepackage{bbm}
\usepackage{geometry}
\usepackage{graphicx}
\usepackage{epstopdf}
\usepackage{graphicx}
\usepackage{enumerate}       
\usepackage[latin1]{inputenc}
\usepackage{graphics,amsthm,amsfonts}
\usepackage{indentfirst}       
\usepackage{amssymb,amsmath,graphicx,color}
\usepackage{comment}
\usepackage{appendix}
\usepackage{array}
\usepackage{multirow}
\usepackage{appendix}
\usepackage{chngcntr} 
\usepackage{float}
\usepackage{setspace}
\usepackage{etoolbox}

\theoremstyle{plain}

\theoremstyle{definition}

\theoremstyle{remark}

\newcommand{\btheta}{\mbox{\boldmath $\theta$}}

\newcommand{\balpha}{\mbox{\boldmath $\alpha$}}
\newcommand{\bbeta}{\mbox{\boldmath $\beta$}}

\newcommand{\bgamma}{\mbox{\boldmath $\gamma$}}

\newcommand{\bepsilon}{\mbox{\boldmath $\epsilon$}}

\newcommand{\bs}{\mbox{\boldmath $s$}}

\newcommand{\bg}{\mbox{\boldmath $g$}}

\begin{document}
\doublespacing
	
	\begin{frontmatter}

	\title{Estimation of underreporting in Brazilian tuberculosis data, 2012-2014}
	\author[1,2]{Guilherme Lopes de Oliveira \corref{mycorrespondingauthor}}
	\address[1]{Departamento de Computa\c c\~ao, Centro Federal de Educa\c c\~ao Tecnol\'ogica de Minas Gerais, Brazil}
	\cortext[mycorrespondingauthor]{Corresponding author. Departamento de Computa\c c\~ao, CEFET-MG, Av. Amazonas, 7675, Nova Gameleira, CEP 30510-000, Belo Horizonte, Minas Gerais, Brazil. E-mail: guilhermeoliveira@cefetmg.br}
		
	\author[2]{Rosangela Helena Loschi}
	\address[2]{Departamento de Estatística, Universidade Federal de Minas Gerais, Brazil}	
%
%
\begin{abstract}
Analysis of burden of underregistration in tuberculosis data in Brazil, from 2012 to 2014. 
Approches of \cite{Oliveira2020} and \cite{Stoner2019} are applied. 
The main focus is to illustrated how the approach of \cite{Oliveira2020} can be applied when the clustering structure is not previously available.
\end{abstract}
%

\end{frontmatter}

\section{Introduction}

Tuberculosis (TB) is one of the world's major public health problems. 
According to the World Health Organization (WHO), TB is the ninth leading cause of death worldwide and the leading cause from a single infectious agent, ranking above HIV/AIDS. In 2016, an estimated 1.7 million people died from TB, including nearly 400,000 people who were co-infected with HIV. 
Brazil is among the top twenty countries by absolute mortality \citep{WHO2017}.
Ending the TB epidemic by 2030 is among the health targets of the Sustainable Development Goals of the United Nations. 
To assess whether these targets are reached and to provide better estimates for the incidence rates, robust monitoring and evaluation of trends in the burden of TB are essential. 

Estimation of TB incidence is a major challenge in many countries due to underreporting and under-diagnosis of TB cases.
Tackling the epidemic requires action to close gaps in care and availability of financial resources. The WHO Global Tuberculosis Report 2017 also evidences that underreporting and underdiagnosis of TB cases continues to be a challenge, especially in countries with large unregulated private sectors and weak health systems \citep{WHO2017}. The WHO has performed some inventory studies to measure the level of TB underreporting in civil registration systems, especially in endemic countries  \citep{WHO2012}.

In Brazil, the Notifiable Diseases Information System (SINAN) provides information about the tuberculosis occurrence, patterns and trends. The  notification of TB cases in SINAN is mandatory and, despite its high spatial coverage, the system is not able to report all TB cases \citep{Stoner2019}.
\cite{Santos2018} showed that the variables associated
with underreporting of TB were mostly related to the healthcare system rather than to individual characteristics of the patients, 	which indicates the need for training the health professionals in order to correctly notify the information in the systems. As pointed out by the Brazilian Ministry of Health \citep{MS2016}, underreporting of TB represents a major loss as it leads to a delay in starting the TB treatment.

In this paper, we apply the model proposed in \cite{Oliveira2020} to estimate the  TB incidence rates in the $A=557$ mainland Brazilian microregions considering SINAN's data from 2012 to 2014. 
We consider usual clustering techniques to define the required data quality groups. That dataset were previously analyzed by \cite{Stoner2019}, from which we obtained all variables considered in our analysis. 
Results are compared to those obtained by fitting  \cite{Stoner2019}'s model.  
It worth mentioning that another interesting modeling framework to account for underreporting is provided in \cite{OlLoAs17} but it will not be considered here.

Based on our final dataset (after usual cleaning for missing data and inconsistent information), we found that between 2012 and 2014 there were 208,901 TB cases notified in the 557  Brazilian microregions. 
As we aim to map the TB incidence in Brazilian territory, we exclude the microregion of \textit{Fernando de Noronha} from our analysis since it is an island with no contiguous neighboring area.

\subsection{Model specification}

We assume $Y_i \mid \theta_i, \epsilon_i  \stackrel{ind}{\sim} {\mathcal{P}oisson}(E_i\theta_i\epsilon_i),~i=1,...,557$, 
where $E_i$ is an offset representing the expected number of events at area $i$ calculated using the naive estimator $ E_i = (N_i \sum_{i = 1}^{557} Y_i) / N $, with $ N_i $ representing the total number of individuals at risk in that area and $N=\sum_{i = 1}^{557} N_i$.
Parameter $\theta_i$ represents the TB relative risk whereas parameter $\epsilon_i$ represents the TB reporting probability or, equivalently, the proportion of the total TB count that is effectively recorded. 

For modeling the relative risks it is assumed a log-linear regression structure  which includes local and spatial random effects, that is, ${\rm log}(\theta_i) = \beta_0 + {\mathbf{X}}_i \bbeta + u_i + s_i,~\forall~i$, 
where $u_i$ and $s_i$ represent the usual local and spatial effects, respectively.
Five covariates are introduced in this regression model: 
the proportion of economically active adults without employment (Unemployment),
the the proportion of people residing in households with more than two persons per room (Density),
the proportion of people living in an urban setting (Urbanisation),
the proportion of the population made up by indigenous groups (Indigenous)
and  average monthly coverage (\%) of the Brazil's Family Health Strategy (ESF)  from 2012 to 2014 in relation to the total population (ESF).

For modeling the reporting probabilities $\bepsilon=(\epsilon_1,...,\epsilon_{557})$ we consider the approach of \cite{Oliveira2020} as detailed in the following. For comparison purpose, we also apply the modeling strategy used by \cite{Stoner2019} to analyze the same TB dataset considered in this work.


The approach for the compound Poisson model proposed in \cite{Oliveira2020} requires the prior specification of data quality groups for the microregions. 
We will refer to such model as the \textit{Clustering Model}.
To define the clustering indicator variable, we performed an usual clustering method with basis on a set of numerical indicators proposed in \cite{Gabriela2017} to evaluate the quality of data recorded in the Brazilian TB surveillance system. 
These authors considered 14 indicators to measure four attributes for the TB data recorded in SINAN from 2012 to 2014: completeness, consistency, timeliness and acceptability. 
More specifically, we collected from  \cite{Gabriela2017} the indicators of \textit{consistency} (percentage of cases with notification date greater or equal to diagnosis date), \textit{completeness} (median for the percentage of completeness measured in five attributes of the SINAN registration form), \textit{timeliness of notification} (percentage of cases with an interval between notification date and diagnosis date smaller or equal to 7 days) and \textit{timeliness of treatment} (percentage of cases with an interval between the date of starting treatment and diagnosis of less than 1 day).  
Besides these four indicators, we included in the clustering analysis the information of two other covariates originated from distinct data sources: the \textit{percentage of general deaths with ill-defined cause} (collected from the Brazilian DATASUS repository for period 2012-2014, available at {http://www2.datasus.gov.br/}) and the \textit{estimated registration coverage for the Brazilian mortality information system} (available from  \cite{SchmertGo18}'s companion website {http://mortality-subregistration.schmert.net/}). Although these two last variables may be more likely related to underreporting of TB deaths rather than TB incidence, we consider they are relevant proxies for general quality of the civil systems for collecting health data in Brazil. As such, they can be helpful in our data quality clustering definition.

The six previous variables was applied to the usual Ward linkage clustering method with the squared Euclidean distance measure. By comparing the similarity measures in the clustering algorithm steps, we found that using $K=23$ groups is an interesting strategy to analyze our TB data in period 2012-2014. 
The groups were labeled into hierarchical data quality categories according to the resulting clusters' centroid (mean). As all variables considered for grouping are measured in an increasing quality scale, we assumed that the greater the cluster mean (centroid), the best the data quality.
The best group (Cluster 1) ended with 32 microregions whereas only 3 microregions were allocated to the worst data quality cluster (Cluster 23).
Then, following \cite{Oliveira2020} we model the TB reporting probabilities in each area $i$ as being
\begin{equation}\label{EqEpsilonTB}
\epsilon_i = 1 - \sum_{j=1}^{23} h_{ji} \gamma_j,
\end{equation}
where $h_{ji}=1$ if area $i$ belongs to cluster $j$ and $h_{ji}=0$ otherwise, for $i=1,\ldots,A$ and $j=1,\ldots,23$. Parameters $\gamma_{1},\ldots,\gamma_{23}$ are related to the clustering underreporting probabilities which are discussed in details in \cite{Oliveira2020}.

For comparison purposes, we also fitted the Brazilian  TB data using the modeling strategy proposed in \cite{Stoner2019}. 
The relative risks $\btheta$ are modeled using the same log-regression structure previously mentioned. \cite{Stoner2019} assumes that the reporting probabilities $\bepsilon$ have the logistic-regression structure given by
\begin{equation}\label{EqEpsilonTBstoner}
{\rm logit}\left(\epsilon_i\right) = \alpha_0 + \bg(w_i)\balpha + \delta_i,~~\mbox{for}~i=1,\ldots,A,
\end{equation}
where $w$ represents the covariate \textit{timeliness of treatment} previously mentioned; $\bg$ is a function defining an orthogonal polynomial of degree 3 introduced to reduce multiple-collinearity and, at the same time, it ensures that $\bg(w)=0$ when $w=\bar{w}$, so that (at the logistic scale) $\alpha_0$ is the mean reporting rate for a region with mean treatment timeliness (for more details, see Section 3 of \cite{Stoner2019}); $\balpha=(\alpha_1, \alpha_2,\alpha_3)$; and $\delta_i$ is a local random effect. We will refer to such model as the \textit{Pogit Model}.

\subsection{About the prior elicitation}

To fit the Clustering Model we adopt the conditional uniform prior distribution given in expression (\ref{UnifPri}) to model the parameter vector $\bgamma=(\gamma_{1},...,\gamma_{23})$, eliciting an informative prior distribution only for parameter $\gamma_1$ (called partially informative prior distribution).
To build such informative prior distribution, we consider available studies on TB underreporting in Brazil. 
As discussed in \cite{Stoner2019}, in 2017 the WHO reported  point estimates for overall TB detection rate in Brazil for years  2012, 2013 and 2014 \citep{WHO2017}. The results are related to inventory study-derived estimates \citep{WHO2012} and revels that, respectively for those years, 91\% (78\%, 100\%), 	84\% (73\%,99\%), and 87\% (75\%,100\%) of TB cases was detected in the Brazilian microregions, where the quantities between parenthesis are the associated 95\% confidence intervals.
%
The reporting probability in those areas experiencing the best data quality (areas within Cluster 1) is likely greater than the overall detection level.
With basis on the findings of previous studies regarding TB underreporting in Brazil \citep{Sousa2011,Sousa2012,Oliveira2012,Gabriela2017, Stoner2019}, we assume that $\gamma_{1}\sim U(0.0,0.05)$ appropriately reflects our prior belief about $\epsilon_i$, for all $i \in \mbox{Cluster 1}$.

When considering the Pogit Model, the prior specification for the reporting probabilities $\bepsilon$ given in equation (\ref{EqEpsilonTBstoner}) are the same considered in \cite{Stoner2019}.
Namely, it is assumed a Gaussian ${\rm{N}}(0,100)$ for the fixed effects $\balpha$ and a Gaussian ${\mathbf{N}}(0, \sigma^2_{\delta})$ for each addictive local effect $\delta_i,~i=1,...,A$. The required informative distributive for parameter $\alpha_0$, which represents the overall mean reporting rate, was assumed to be a Gaussian distribution  ${\rm{N}}(2, 0.36)$ with basis on the information provided by the WHO. It is worth noticing that all Gaussian distributions described in this section are parametrized in terms of mean and variance. 	

Regarding the prior specification for the log-linear structure imposed to the relative risks $\btheta$, in both models we assume mean-centered covariates with Gaussian prior ${\rm{N}}(0,100)$ for their fixed effects $\bbeta=(\beta_1,\dots,\beta_5)$. 
Following \cite{Stoner2019}, we assume \textit{a priori} that $\beta_0\sim {\rm{N}}(-8, 1)$. This was specified by those authours using a prior  predictive checking and it reflects the belief that very high values (such as over 1 million) for the total number of TB cases are unlikely.
Additionally,  we assume that $u_i \buildrel iid \over\sim {\rm{N}}(0, \sigma^2_u)$ and that $\bs=(s_1, \dots, s_{A})$ have the ICAR prior distribution \citep{BeYoMo91} with precision parameter $\tau_s=\sigma^{-2}_s$. 
Following  \cite{Stoner2019}, the prior distributions for variances $\sigma^2_u$, $\sigma^2_s$ and $\sigma^2_{\delta}$ are truncated ${\rm{N}}(0, 1)$ with domain restrict to $(0,\infty)$. Such a choice reflects the belief that low variance values are more likely than higher ones.

The MCMC scheme is performed using package Nimble \citep{deValpire2017} from software R \citep{softwareR}. The basic script for running each model is available in the supplementary materials of \cite{Stoner2019} and \cite{Oliveira2020}. 
For both models two chains were considered each with a total of 3,000,000 iterations, being the first 1,000,000  discarded as a burn-in period and a lag of 3,000 iterations was selected in order to avoid autocorrelated posterior samples. 
Trace plots for the MCMC samples were inspected and the potential scale reduction factor (PSRF) \citep{BrooksGelman1998} was calculated as less than 1.04 and 1.06 for all regression coefficients and variance parameters, respectively, in the Clustering and the Pogit Models, thus indicating convergence.

%


%


\subsection{Posterior Results}

\begin{figure}[htb!]
	\subfigure[$\hat{\btheta}$: Clustering Model with $K=23$]{
		\includegraphics[scale=0.4]{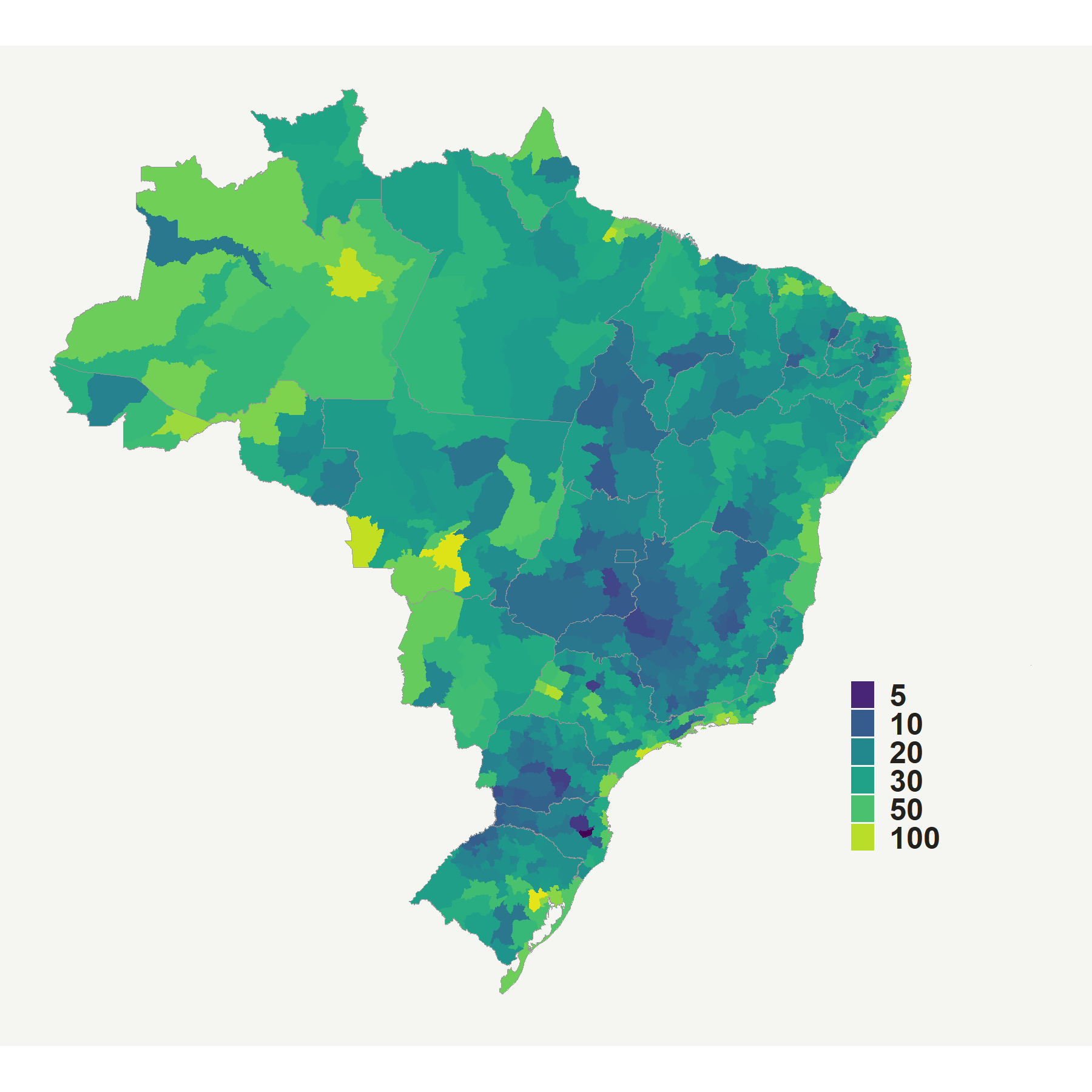}
	}
	\subfigure[$\hat{\bepsilon}$: Clustering Model with $K=23$]{
		\includegraphics[scale=0.4]{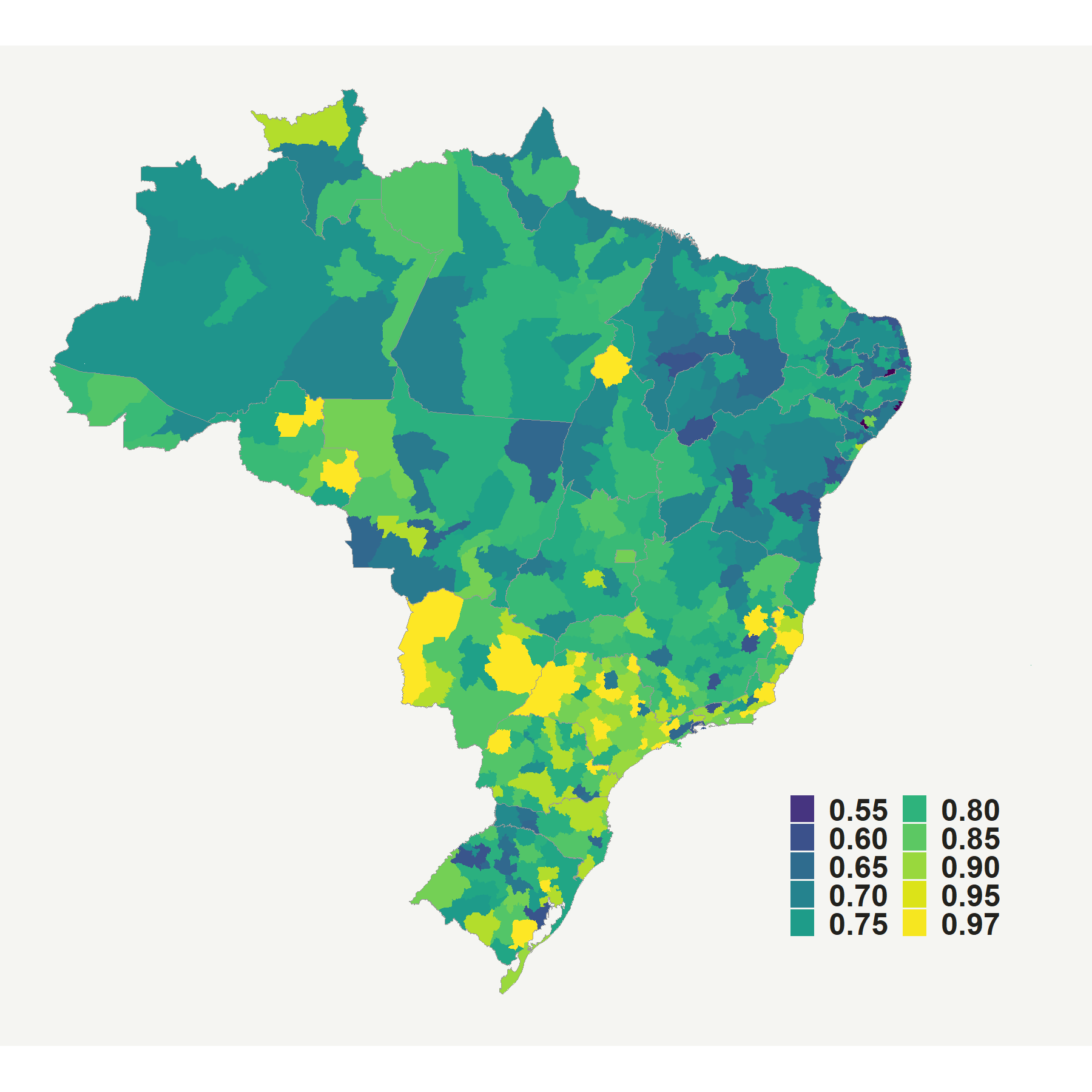}
	} \\
	\subfigure[$\hat{\btheta}$: Pogit Model]{
		\includegraphics[scale=0.4]{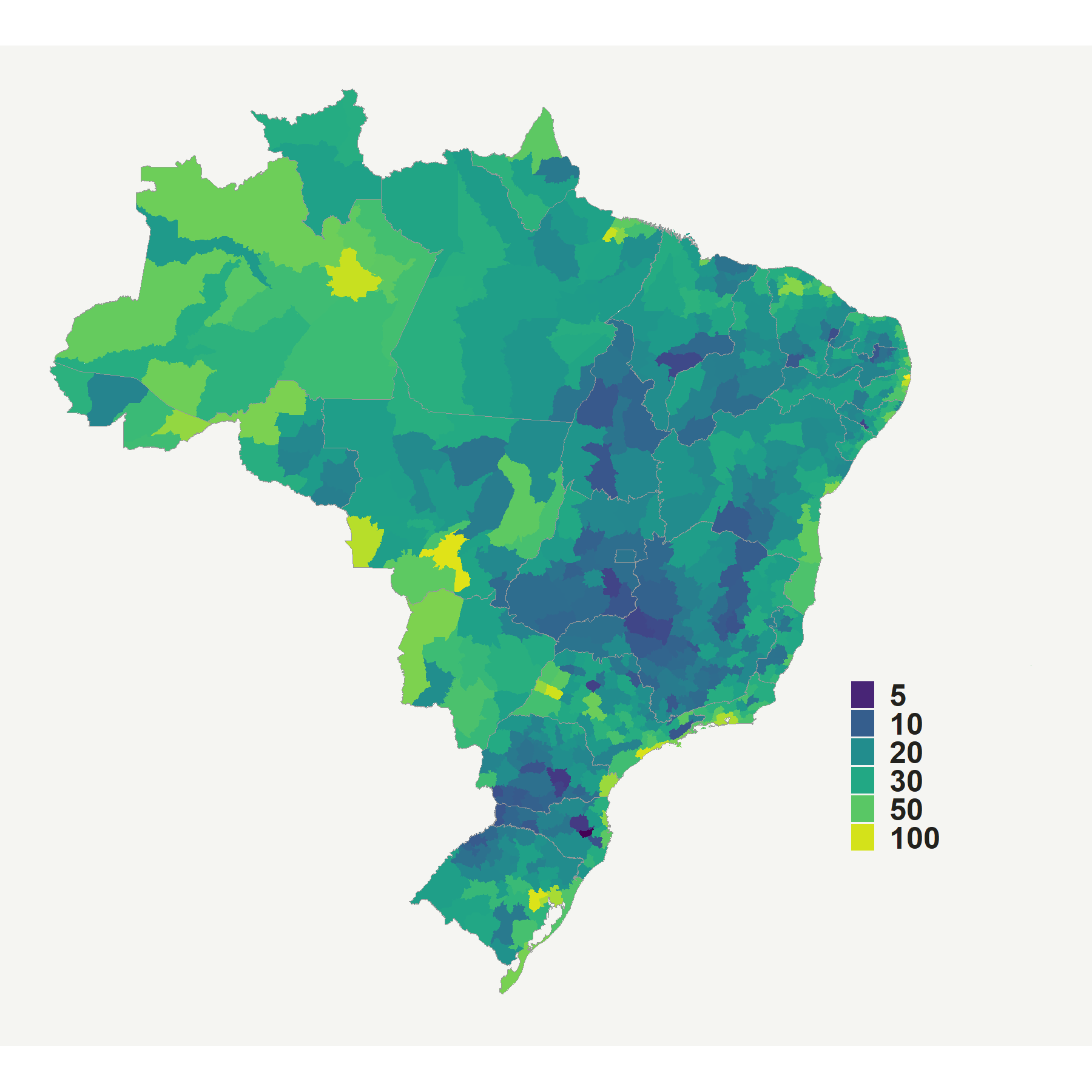}
	}
	\subfigure[$\hat{\bepsilon}$: Pogit Model]{
		\includegraphics[scale=0.4]{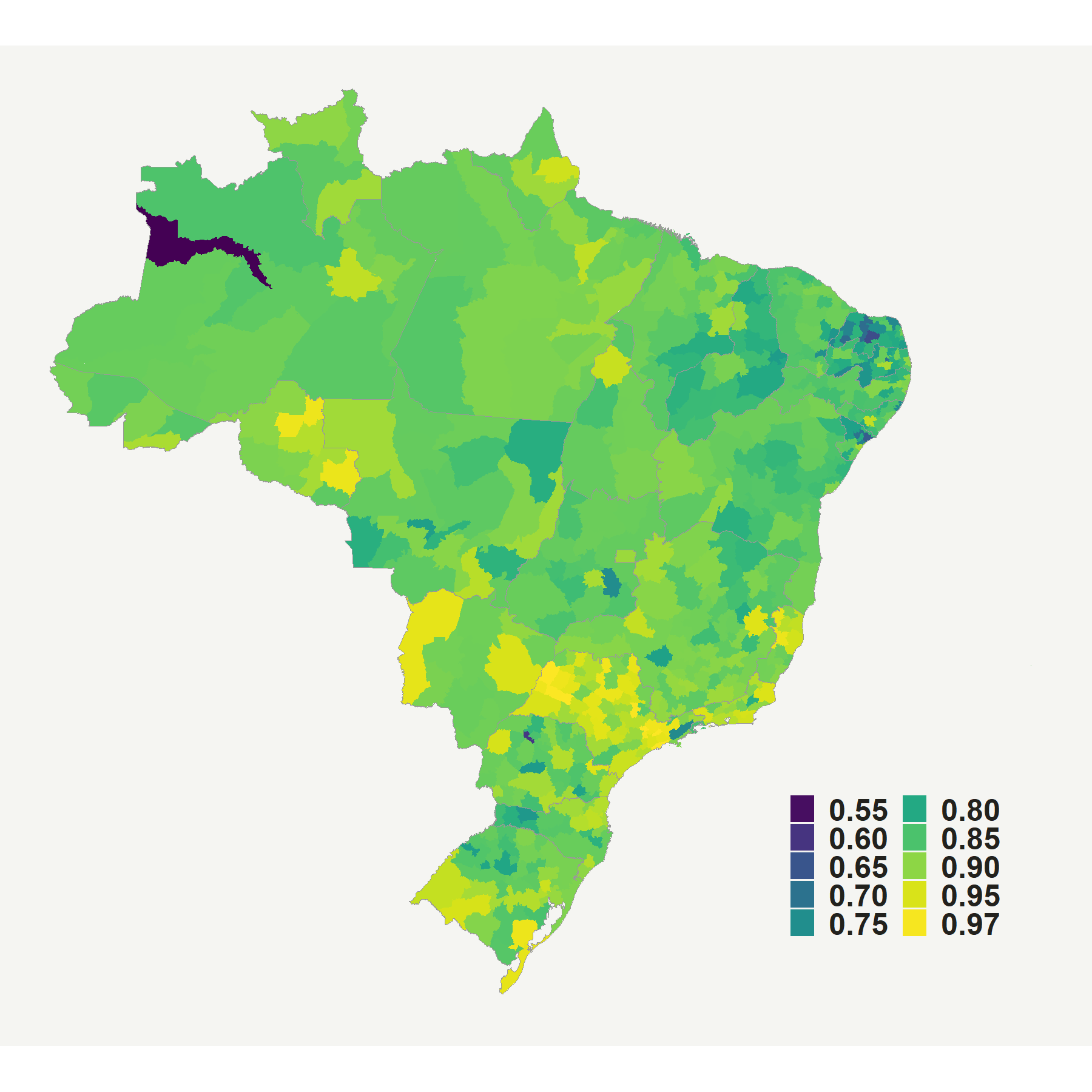}
	}
	\caption{Posterior mean for the tuberculosis incidence rates per 100,000 inhabitants (left) and the reporting probability (right) under the model proposed in \cite{Oliveira2020} with $K=23$ data quality clusters (top), denoted by Clustering Model, and also under the modeling strategy proposed in \cite{Stoner2019} (bottom), denoted by Pogit Model; Brazilian data, 2012-2014.}
	\label{figTBdata2012_2014}
\end{figure}

Figure \ref{figTBdata2012_2014} displays the spatial distribution of the estimated TB incidence rates per 100,000 inhabitants and the respective reporting probabilities throughout the $557$ mainland Brazilian microregions under the Censoring Model and the Pogit Model. 
The models provided a quite similar spatial structure for the tuberculosis incidence (Panels (a) and (c)), with highest values mainly concentrated in the North and Central-West regions of the country. Clusters of microregions with elevated values for the disease incidence can also be observed along the coast of Brazil (right edge of the map). 

Regarding to the reporting probabilities, from Panels (b) and (d) of Figure \ref{figTBdata2012_2014}, it can be seen that, except for an specific microregion in the Northwest of Brazil, both models provide a similar spatial pattern. Although the reporting probabilities showed to be more homogeneous under the Pogit model, there is an agreement in relation to the clusters of areas with the highest and smallest values for the posterior estimates of $\bepsilon$, specially regarding the highest values. In general, estimates under the  Pogit Model are more concentrated in greater values than those observed under the Clustering Model.

The minimum estimate for the reporting probabilities under the Pogit Model is 0.5352, the only estimate smaller than 0.60. Under such a model, the first quartile, mean, third quartile and maximum estimates were, respectively, 0.8515, 0.8722, 0.8978 and  0.9748. The four smallest estimates (all above 0.675) were observed in microregions in which the value for covariate $w$ is zero. Furthermore, these four microregions figured out among the 10\% smallest populations and 17\% lowest TB counts. 
This analysis suggests that the estimation of the reporting probabilities is highly influenced by the reporting proxy variable considered in the logistic regression. 
In small populations, the variable \textit{timeliness of treatment} ($w$) may not be measured properly, inducing to discrepant results.

A similar analysis under the Clustering model revels that the minimum estimate for the reporting probabilities is 0.4756 (within Cluster 23). This is the only cluster with an estimate for the reporting probability smaller than 0.60 and it composed by 3 microregions. 
Under such a model, the first quartile, mean, third quartile and maximum estimates were, respectively, 0.7128, 0.7865, 0.8416 and  0.9759. The three regions within the worst cluster figured out among the 13\% smaller populations and 12\% lower TB counts. The quite discrepant small value for the reporting probability in this cluster may be related to the fact that it only contains  microregion with small populations.

Table \ref{tabTB} summarizes the results for relative risks $\btheta$.
In both models,  the  95\% highest posterior density interval ($HPD_{95\%}$) 
for covariate \textit{ESF} contains the value zero, thus indicating that this proxy for access to healthcare does not have a significant (non-null) effect in the tuberculosis incidence rate. 
Among the other covariates, only the effect of \textit{Density} was estimated differently under the two fitted models. Result provided by the Clustering Model is more consistent with what is expected in practice for the effect of such covariate.

The log pseudo-marginal likelihood (LPML) criterion \citep*{Ibrahim01} points that the TB data is better fitted by the Clustering Model.
Such model is more parsimonious than the Pogit Model since only $K=23$ parameters are	estimated in the reporting mechanism instead of estimating the fixed effects $\alpha_0$ and $\balpha$ besides the local effects $\delta_1,\ldots,\delta_{557}$. 
In some sense, there are more data information available to estimate each unknown parameter associated to $\bepsilon$ under the Clustering Model than under the Pogit Model. This might be one of the reasons for the slightly better performance of the former model in relation to the latter.

\begin{table}[htb!]
	\centering
	\caption{Posterior summaries for the regression effects $\beta_0$ and $\bbeta$ under the model proposed in \cite{Oliveira2020} with $K=23$ data quality clusters, denoted by Clustering Model, and also under the modeling strategy proposed in \cite{Stoner2019}, denoted by Pogit Model; Brazilian tuberculosis data for period 2012-2014. We provide the posterior mean (Mean), the posterior standard deviation (St.Dev.) and the 95\% highest posterior density interval ($HPD_{95\%}$).}\vspace{0.3cm}
	\tabcolsep=4.5pt 
	\begin{tabular}{cccccccc}
		\hline
		Covariate    & Mean & St.Dev. & $HPD_{95\%}$ &  &.Mean & St.Dev. & $HPD_{95\%}$\\
		\hline
		& \multicolumn{3}{c}{Clustering Model (\textbf{LPML=-2505.357})} 	& & \multicolumn{3}{c}{Pogit Model (LPML=-2528.662)} \\ 
		\cline{2-4} \cline{6-8}
		Intercept   &-8.254 &0.052  & (-8.353,-8.153)  &  &-8.360 &0.065  & (-8.468,-8.245)   \\
		Unemployment      &0.119 &0.027  & (0.064,0.169)   &  &0.047 &0.011  & (0.026,0.068)   \\
		Density      &0.125 &0.044  & (0.043,0.212)   & &-0.218 &0.009  & (0.003,0.016) \\
		Urbanisation      &0.206 &0.030   & (0.152,0.268)   & &0.014 &0.002   & (0.010,0.017)  \\
		Indigenous   &0.056 &0.018   & (0.019,0.090)   & & 0.015 &0.005   & (0.005,0.025)  \\
		ESF 		&-0.026 &0.025   & (-0.078,0.020)   &  &-0.001 &0.001   & (-0.003,0.001)   \\
		\hline
		\hline     
	\end{tabular}
	\label{tabTB}
\end{table}

\section{Discussion}

We addressed an important problem in Epidemiology and public health fields.
Providing realistic estimates for the TB incidence rates is important to guide healthcare professionals in making their decisions to control the endemic disease.

The correction of underreporting bias in Brazilian TB counts, 2012-2014, was performed using the approaches introduced by \cite{Oliveira2020} and \cite{Stoner2019}. 
They provided a quite similar spatial pattern for the disease incidence rates.
For the reporting probabilities, estimates under the Clustering Model \citep{Oliveira2020}  showed a greater discrepancy throughout the country if compared to the ones obtained under the Pogit Model \citep{Stoner2019}. 

For the Clustering Model, information from six data quality indicators was taken into consideration to define the groups, including that one used as TB reporting proxy in the logistic regression assumed for the Pogit Model. 
The effort to define the grouping was compensated by a better data fitting, according to the LPML measure. It worth noting, however, that the findings of this applied analysis cannot be generalized to other examples without further exhaustive investigation.

It is intended to make a more robust comparison between the methods through simulated scenarios. In the TB data analysis, it is also of interest to perform a sensitivity analysis regarding effects of different clustering definitions under the \cite{Oliveira2020}'s approach. Likewise, we aim to fit the model of \cite{Stoner2019} using different proxies in the logistic regression. Some discussions on these regards are presented in the cited papers but we intend to do additional studies focusing on this specific application and possibly others.

\section*{Acknowledgements}
G. L. de Oliveira , R. H. Loschi and Renato M. Assun\c c\~ao thank the Brazilian funding aggencies CNPq,  CAPES ({\it Coordena\c c\~ao de Aperfei\c coamento de Pessoal Coordenação de Aperfeiçoamento de Pessoal de N\'{\i}vel Superior}) , and FAPEMIG ({\it Funda\c c\~ao de Amparo \`a Pesquisa do Estado de Minas Gerais})  for partially support their research.
The authors also thank Gabriela Drummond Marques da Silva for supporting with the datasets and discussion about the prior specifications in the fitted models.

\section*{Conflicts of Interest}
None declared.



\begin{thebibliography}{10}
	\providecommand{\MR}{\relax\unskip\space MR }
	\providecommand{\url}[1]{\normalfont{#1}}
	\providecommand{\urlprefix}{Available at }
	
	
	\bibitem[\protect\citeauthoryear{\nobreak Besag, \nobreak York,  and Molli\'e}{Besag \textit{et al.}}{1991}]{BeYoMo91} Besag, J., York, J., and Molli\'e, A. (1991). Bayesian image restoration, with two applications in spatial statistics. {\it Annals of the Institute of Statistical Mathematics}, {\bf 43}(1), 1--20.
	
	\bibitem[\protect\citeauthoryear{Ministério da Saúde do Brasil}{2016}]{MS2016}
	Brasil. Ministério da Saúde. Programa Nacional de Controle da Tuberculose [Internet]. Brasília: Ministério da Saúde; 2016 [citado 23 abr. 2017]. \urlprefix\url{http://portalarquivos.saude.gov.br/images/pdf/2017/
		fevereiro/21/Apresentacao-sobre-os-principaisindicadores-	da-tuberculose.pdf}.
	
	
	\bibitem[\protect\citeauthoryear{Brooks and Gelman}{1998}]{BrooksGelman1998} Brooks, S.P. and Gelman, A. (1998). General Methods for Monitoring Convergence of Iterative Simulations. {\it Journal of Computational and Graphical Statistics}, \textbf{7}, 434--455.
	
	\bibitem[\protect\citeauthoryear{de Valpire \textit{et al.}}{2017}]{deValpire2017}
	de Valpine, P., Turek, D., Paciorek, C.J., Anderson-Bergman, C., Lang, D.T.
	and Bodik, R. (2017). Programming With Models: Writing Statistical Algorithms for General Model Structures With Nimble. {\it Journal of Computational and Graphical Statistics}, \textbf{26}, 403--413.
	
	\bibitem[\protect\citeauthoryear{\nobreak Ibrahim, \nobreak Chen,  and Sinha}{Ibrahim et. al}{2001}]{Ibrahim01}
	Ibrahim, J. G., Chen, M-H.,  and Sinha, D. (2001) {\it Bayesian Survival Analysis}. New York: Springer-Verlag; 2001. pp. 589.
	
	\bibitem[\protect\citeauthoryear{Oliveira \textit{et al.}}{2020}]{Oliveira2020}
	Oliveira, G.L., Argiento, R., Loschi, R.H., Assun\c c\~ao, R.M., Ruggeri, F. and Branco, M.D. (2020). Bias correction in clustered underreported data. {\it Bayesian Analysis}, Advance Publication, 1--32. DOI: 10.1214/20-BA1244.
	
	\bibitem[Oliveira \textit{et al.} (2017)]{OlLoAs17}
	Oliveira, G.L., Loschi, R.H. and Assun\c c\~ao, R.M. (2017).
	A random-censoring Poisson model for underreported data.
	{\em{Statistics in Medicine}}, {\bf{36}}(30), 4873--4892.
	
	\bibitem[\protect\citeauthoryear{Oliveira \textit{et al.}}{2012}]{Oliveira2012}
	Oliveira, G.P., Pinheiro, R.S., Coeli, C.M., Barreira, D. and Codenotti, S.B. (2012). Mortality information system for identifying underreported cases of tuberculosis in Brazil. {\it Revista Brasileira de Epidemiologia}, \textbf{15}(3), 468--477.
	
	
	\bibitem[\protect\citeauthoryear{R Core Team }{2015}]{softwareR} R Core Team (2015). { R: A Language and Environment for Statistical Computing}, R Foundation for Statistical Computing, Vienna, Austria.  {ISBN} 3-900051-07-0, (2015). \urlprefix\url{https://www.R-project.org/}.
	
	
	\bibitem[\protect\citeauthoryear{Santos \textit{et al.}}{2018}]{Santos2018} Santos, M.L, Coeli, C.M., Batista, J.L., Braga, M.C., de Albuquerque, M.F.P.M. (2018). Factors associated with underreporting of tuberculosis based on data from Sinan Aids and Sinan TB. {\it Revista  Brasileira de Epidemiologia} [online], {\bf{21}}:(e180019).
	
	
	\bibitem[\protect\citeauthoryear{Schmertmann and Gonzaga}{2018}]{SchmertGo18} Schmertmann, C. and Gonzaga, M. R. (2018). Bayesian estimation of age-specific mortality and life expectancy for small areas with defective vital records. {\it{Demography}}, {\bf 55}(4), 1363--1388.
	
	\bibitem[\protect\citeauthoryear{Silva \textit{et al.}}{2017}]{Gabriela2017}
	Silva, G.D.M., Bartholomay, L., Cruz, O.G. and Garcia, L.P. (2017). Evaluation of data quality, timeliness and acceptability of the tuberculosis surveillance system in Brazil's micro-regions. {\it Ciência \& Saúde Coletiva} [online], {\bf{22}}(10), 3307--3319.
	
	
	\bibitem[\protect\citeauthoryear{Sousa and Pinheiro}{2011}]{Sousa2011}
	Sousa, L.M., Pinheiro, R.S. (2011). Unnotified deaths and hospital
	admissions for tuberculosis in the municipality of Rio de Janeiro. 
	{\it Revista de Saúde Pública}, \textbf{45}(1), 31--39.
	
	\bibitem[\protect\citeauthoryear{Sousa \textit{et al.}}{2012}]{Sousa2012}
	Sousa, M.G.G., Andrade, J.R.S., Dantas, C.F. and Cardoso, M.D. (2012).
	Investigação de óbitos por tuberculose, ocorridos na Região Metropolitana do Recife (PE), registrados no Sistema de Informação de Mortalidade, entre 2001 e 2008 (in Portuguese). {\it Cadernos Saúde Coletiva}, \textbf{20}(2), 153--60.
	
	\bibitem[\protect\citeauthoryear{\nobreak Stoner, \nobreak Economou,  and Drummond}{Stoner \textit{et al.}}{2019}]{Stoner2019} Stoner, O; Economou, T; Drummond, G. (2019). A Hierarchical Framework for Correcting Under-Reporting in Count Data. {\it Journal of the American Statistical Association}, {\bf{114}}(528), 1481--1492.
	
	\bibitem[\protect\citeauthoryear{World Health Organization}{2012}]{WHO2012} World Health Organization (WHO). (2012). Assessing tuberculosis under-reporting through inventory studies.
	{ \it WHO Library Cataloguing-in-Publication Data}. France, WHO/HTM/TB/2012.12, ISBN 978-92-4-150494-2.
	
	
	\bibitem[\protect\citeauthoryear{World Health Organization}{2017}]{WHO2017} World Health Organization (WHO). (2017). Global Tuberculosis Report 2017.
	{ \it WHO Library Cataloguing-in-Publication Data}. Switzerland, WHO/HTM/TB/2017.23, ISBN 978-92-4-156551-6.
	
\end{thebibliography}

\end{document}